\documentclass[pra,aps,10pt,twocolumn,floatfix,nofootinbib]{revtex4-1}
\usepackage{geometry}
\usepackage{graphicx}
\usepackage{amsfonts}
\usepackage{amsmath}
\usepackage{bm}
\usepackage{url}
\usepackage{color}
\usepackage{xfrac}
\usepackage{array,url,kantlipsum}
\usepackage[braket,qm]{qcircuit}
\usepackage{physics}
\usepackage{mathtools}
\usepackage[dvipsnames]{xcolor}

\usepackage{hyperref}
\hypersetup{%
  colorlinks=true,
  allcolors=Blue
 }

\DeclarePairedDelimiter{\ceil}{\lceil}{\rceil}

\begin{document}
\title{Robust readout of bosonic qubits in the dispersive coupling regime} 

\author{Connor~T.~Hann}
\author{Salvatore S.~Elder}
\author{Christopher~S.~Wang}
\author{Kevin~Chou}
\author{Robert~J.~Schoelkopf}
\author{Liang~Jiang}
\affiliation{\mbox{Departments of Applied Physics and Physics, Yale University, New Haven, Connecticut 06511, USA}}
\affiliation{Yale Quantum Institute, Yale University, New Haven, Connecticut 06511, USA }

\begin{abstract}
High-fidelity qubit measurements play a crucial role in quantum computation, communication, and metrology. In recent experiments, it has been shown that readout fidelity can be improved by performing repeated quantum non-demolition (QND) readouts of a qubit's state through an ancilla. For a qubit encoded in a two-level system, the fidelity of such schemes is limited by the fact that a single error can destroy the information in the qubit. On the other hand, if a bosonic system is used, this fundamental limit can be overcome by utilizing higher levels such that a single error still leaves states distinguishable. 
In this work, we present a robust readout scheme which leverages both repeated QND readouts and higher-level encodings to asymptotically suppress the effects of mode relaxation and individual measurement infidelity. We calculate the measurement fidelity in terms of general experimental parameters, provide an information-theoretic description of the scheme, and describe its application to several encodings, including cat and binomial codes.
\end{abstract}

\maketitle 


\section{Introduction}

The ability to measure a qubit with high fidelity is of great importance in quantum computation \cite{Divincenzo2000,Knill2005} 
and metrology \cite{Blatt2008,Giovannetti2011}, 
as well as in measurement-based feedback control \cite{Vijay2012,Sayrin2011,Cook2007,Yamamoto2008,Cramer2016,Ofek2016} 
and computation \cite{Raussendorf2003,Gross2007,Briegel2009}.
Experimentally, much progress has been made in recent years toward realizing high-fidelity qubit measurement.  High-fidelity single-shot measurements have been demonstrated in a wide variety of physical systems, including nitrogen-vacancy centers \cite{Robledo2011,Shields2015,DAnjou2016}, superconducting circuits \cite{Reed2010,Jeffrey2014,Walter2017}, and quantum dots \cite{Barthel2009,Harvey-Collardnodate,Nakajima2017}. Qubit relaxation is often a limiting factor in such experiments, and in systems with longer qubit lifetimes higher readout fidelities are possible. Indeed, in trapped ions---known for their long coherence times---readout fidelities in excess of $99.9\%$~\cite{Noek2013,Harty2014} and even $99.99\%$~\cite{Myerson2008,Burrell2010} have been demonstrated experimentally. 

\begin{figure}[bp]
\begin{center}
\includegraphics[width=\columnwidth]{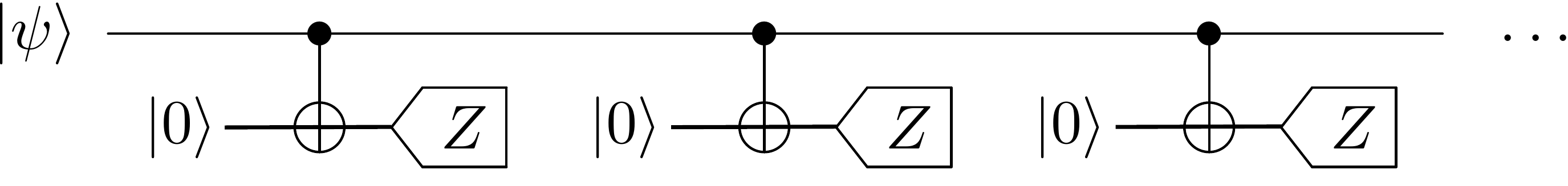}
\caption{Repeated QND readouts. Contributions to overall measurement infidelity from ancilla preparation/readout errors and noise can be exponentially suppressed by repeating measurements. }
\label{fig:QNDcircuit}
\end{center}
\end{figure}

While this experimental progress is encouraging, strategies to further improve qubit readout fidelity are of great interest. One such strategy involves coupling the primary qubit to an ancillary readout qubit. Measurements are performed by mapping the system's state onto the ancilla, whose state is then read out. These measurements are said to be quantum non-demolition (QND) if the system's measurement eigenstates are unaffected by the ancilla readout procedure. QND measurements are necessarily repeatable, and the overall measurement fidelity can be improved by repeating measurements to suppress individual measurement infidelity (\hyperref[fig:QNDcircuit]{Fig.~\ref{fig:QNDcircuit}}).  Highly QND readouts have already been realized in trapped-ion systems \cite{Hume2007,Wolf2016}, nitrogen vacancy centers \cite{Jiang2009,Neumann2010,Cramer2016}, and circuit QED systems~\cite{Johnson2010, Peaudecerf2014,Sun2014, Saira2014,Ofek2016,Lupascu2007, Vijay2011}.

For a qubit encoded in a two-level system, the fidelity of such repeated readout procedures is fundamentally limited by the fact that there exist single errors, such as relaxation of the excited state to the ground state, that can destroy the information in the qubit. This fundamental limit can be overcome, however, by robustly encoding the information within a larger Hilbert space, so that single errors leave states distinguishable.
The combination of repeated QND measurements and robust encoding thus enables one to overcome limits imposed by both individual measurement infidelity and qubit relaxation. 

In this work we propose a robust readout scheme for bosonic systems in the dispersive coupling regime---a class of systems where information can be both encoded robustly and read out in a QND way. The infinite-dimensional Hilbert space of a single bosonic mode (quantum harmonic oscillator) provides room to encode information and protect it from errors~\cite{Cochrane1999, Gottesman2001, Michael2016}, while the mode's dispersive coupling to an ancillary quantum system enables repeated QND readout~\cite{Brune1992,Blais2004,Wallraff2004,Wallraff2005,Sun2014}. We show explicitly how the combination of these two techniques allows one to simultaneously suppress the contributions to readout infidelity from qubit relaxation and individual measurement noise to higher order, potentially yielding orders-of-magnitude improvement in readout fidelity. 


This scheme is applicable to a variety of systems where bosonic modes, typically in the form of photons or phonons, are naturally available. Circuit QED systems, where the strong dispersive regime is experimentally accessible~\cite{Wallraff2004,Schuster2007,Boissonneault2009}, provide one example. Other examples include optomechanical systems, where dispersive couplings necessary for QND readout have been demonstrated~\cite{Jayich2008, Thompson2008}, and in principle also nanomechanical systems~\cite{LaHaye2009, OConnell2010} or circuit quantum acoustodynamic systems~\cite{Manenti2017,Chu2017}, provided sufficiently strong couplings and long mode lifetimes can be engineered. 
More broadly, this scheme can be applied in any system where a bosonic mode has a strong dispersive coupling to an ancilla, so it can even be applied to more exotic systems, e.g.~quantum magnonics, where strong dispersive couplings were recently demonstrated~\cite{Lachance-Quirion2017}.


This article is organized as follows. In \hyperref[decay-twolevel]{Sec.~\ref{decay-twolevel}}, we use a simple Fock state encoding to introduce the robust readout scheme for a lossy bosonic mode dispersively coupled to a two-level readout ancilla. This encoding serves as a straightforward example of an encoding suited to robust readout, and its analysis (Secs.~II-V) is intended to make the ideas underlying the robust readout scheme abundantly clear. With this encoding, we explicitly compute the readout infidelity and show that contributions from relaxation and individual measurement noise are suppressed. 
In \hyperref[decayexcite-twolevel]{Sec.~\ref{decayexcite-twolevel}} we generalize the readout scheme so that contributions to the infidelity from spontaneous heating are also suppressed.
In \hyperref[decayexcite-manylevel]{Sec.~\ref{decayexcite-manylevel}} we show how, given a readout ancilla with more than two levels, readout fidelity can be significantly improved by using a maximum likelihood estimate as opposed to simple majority voting. 
In \hyperref[InfoTheory]{Sec.~\ref{InfoTheory}}, we consider the robust readout scheme from the perspective of classical information theory and place a lower bound on readout infidelity. This concludes our analysis of the Fock state encoding. 
We consider other encodings in \hyperref[Encodings]{Sec.~\ref{Encodings}}, which contains the main results of this article. We identify  criteria on encodings that are sufficient for robust, ancilla-assisted readout of a qubit encoded in a lossy bosonic mode, and as examples, 
we explicitly show that these criteria are satisfied by cat codes and binomial codes. We approximate the readout fidelity for both codes. 


\section{Robust readout of a qubit encoded in a lossy bosonic mode }
\label{decay-twolevel}
\subsection{Robust readout scheme}

Let a bit of quantum information be encoded in a bosonic mode as $\ket{\psi}_B = \alpha \ket{0}_B + \beta \ket{1}_B$, where $\ket{0}_B$ and $\ket{1}_B$ are the ``logical'' states in the mode's Hilbert space that we seek to distinguish with maximal fidelity. Readout of this qubit (henceforth referred to as the bosonic qubit) is performed by repeatedly mapping its state onto a two-level ancillary quantum system, whose state is subsequently measured. The mapping and ancilla readout processes are assumed to be QND so that they can be repeated without disturbing the bosonic qubit. This assumption is justified in the dispersive coupling regime, as will be shown explicitly.  

\begin{figure}[htbp]
\begin{center}
\includegraphics[width=\columnwidth]{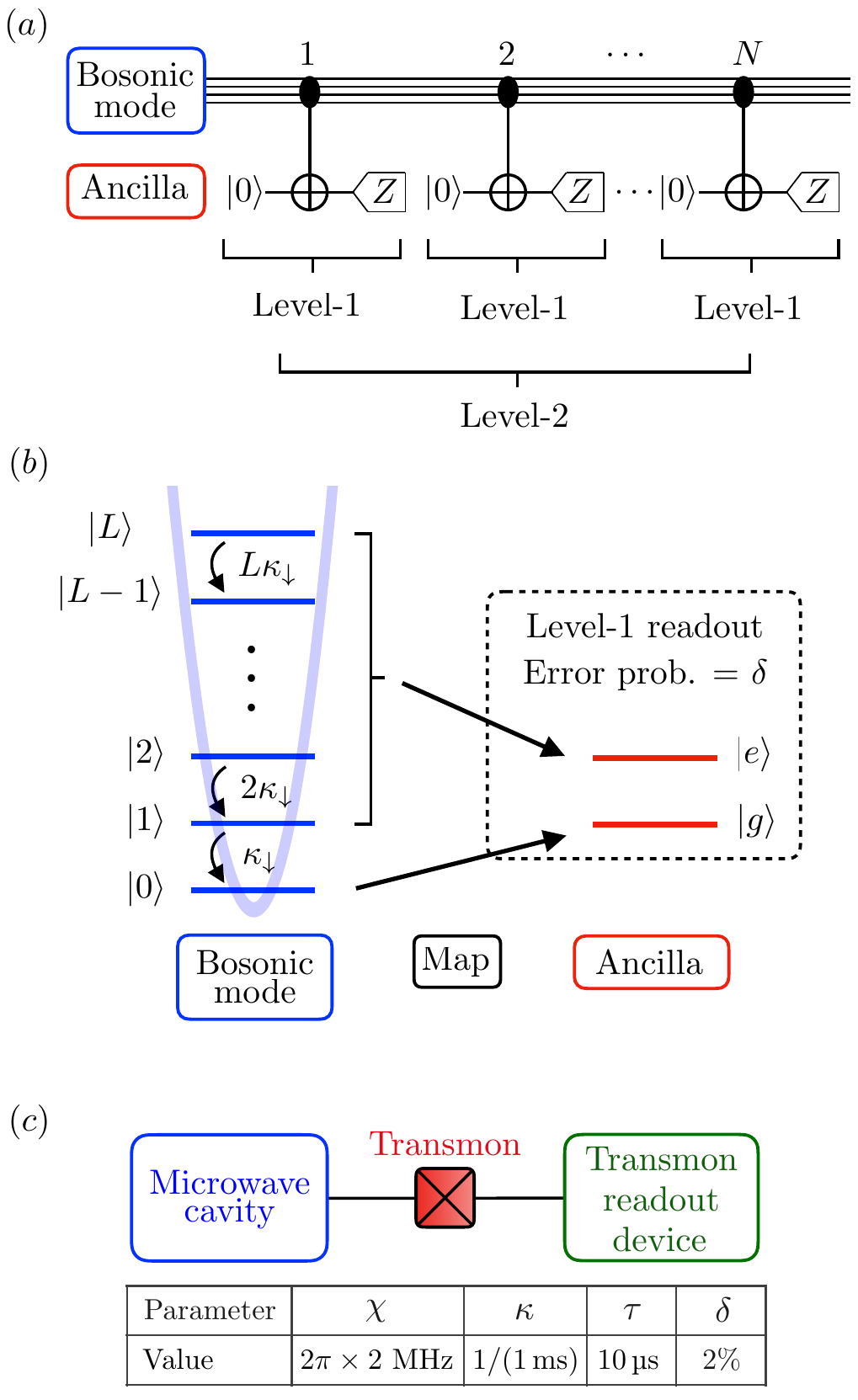}
\caption{Robust readout scheme. (a) Quantum circuit for the readout scheme. The state of the bosonic qubit is read out through repeated QND mappings of its state onto an ancilla. (b) The bosonic mode and mapping procedure. Fock states in the bosonic mode decay with rates proportional to their excitation number. All excited states are mapped to the excited state of the two-level readout ancilla. (c) Schematic of a circuit QED system. The state of a microwave cavity mode can be read out via a dispersive coupling to a transmon qubit. The transmon is measured via its coupling to some other device, typically a resonator or cavity. Realistic parameter values for this architecture, c.f.~Refs.~\cite{Heeres2017,Chou2018,Axline2018}, are shown in the table (see \hyperref[decay-twolevelB]{Sec.~\ref{decay-twolevelB}} for parameter definitions). }
\label{fig:setup}
\end{center}
\end{figure}

We refer to the process of mapping the bosonic qubit onto the ancilla, followed by ancilla readout, as a level-1 readout. Each level-1 readout yields one classical bit of information (the ancilla is either found to be in $\ket{g}$ or $\ket{e}$). In our scheme, $N$ repeated level-1 readouts are performed, and their outcomes are collectively analyzed, e.g.~with majority voting, to yield a single bit of classical information (the bosonic qubit is determined to be in either $\ket{0}_B$ or $\ket{1}_B$). We refer to the entire procedure---performing $N$ level-1 readouts and combining the results---as a level-2 readout. This scheme is shown schematically in \hyperref[fig:setup]{Fig.~\ref{fig:setup}(a)}.
 
We now define the logical states, the specific mapping required for this scheme, and the relaxation properties of the bosonic mode, all of which are summarized in \hyperref[fig:setup]{Fig.~\ref{fig:setup}(b)}. The logical states encoding the bosonic qubit are chosen to be the Fock states
 \begin{align}
 \label{Fock_logical_states}
 \ket{0}_B &= \ket{0} \nonumber \\
 \ket{1}_B &= \ket{L}, 
 \end{align}
for positive integer $L$.
We make three remarks on this choice of encoding. First, the reason that we begin by considering this ``Fock code'' is that the analysis of its readout fidelity is straightforward, so the code serves as an instructive example. Second, we note that the Fock code has previously been used in quantum information processing applications. For example, the initialization~\cite{Heeres2015, Heeres2017, Chou2018,Axline2018,Hofheinz2008,Hofheinz2009} and manipulation~\cite{Chou2018} of qubits with this encoding have been demonstrated experimentally. Third, the Fock code is a quantum error-detecting code, capable of detecting excitation loss errors for $L>1$. Thus, this code could be useful in a concatenated encoding scheme, for example, since the ability to detect errors at one level enables more efficient correction of errors at the next level of encoding~\cite{Knill2005}. Other possible choices of the logical states, including quantum error-correcting codes like the cat and binomial codes, are considered in \hyperref[Encodings]{Sec.~\ref{Encodings}}.


In the dispersive coupling regime, the logical states (\hyperref[Fock_logical_states]{\ref{Fock_logical_states}}) can be distinguished through a measurement procedure that is QND. In this work we consider projective measurements and define QND as follows. A projective measurement can be described by a collection of measurement operators $\{\hat{M}_k\}$ that constitute a complete set of orthogonal projectors, satisfying $\hat{M}_k^2 = \hat{M}_k$ and $\sum_k \hat{M}_k = 1$. Such a measurement is QND if 
\begin{equation}
\label{qnd_criterion}
\left[\hat{M}_k, \hat H(t) \right] = 0,
\end{equation}
for all $k$ and $t$, where $\hat{H}(t)$ is an operator describing the ancilla preparation, its coupling to the bosonic mode, and the ancilla readout. In the robust readout scheme, the level-1 measurements are defined by operators $\hat M_0$ and $\hat M_1$ that act on the Hilbert space of the bosonic mode
\begin{align}
\label{Eqn:decay_meas_ops}
\hat{M}_0 &= \ket{0}\bra{0} \nonumber \\
\hat{M}_1 &= \ket{1}\bra{1} + \ldots + \ket{L}\bra{L}.
\end{align}
For a bosonic mode dispersively coupled to a two-level ancilla, QND measurements are possible because these operators commute with the dispersive coupling Hamiltonian, 
\begin{equation}
H_{DC}/\hbar = -\chi \,\hat{a}^\dagger \hat{a} \ket{e}\bra{e},
\end{equation}
where $\ket{g}$ and $\ket{e}$ denote the basis states of the ancilla, and $\hat a$ is the bosonic annihilation operator. Similar QND measurements have already been demonstrated experimentally in circuit QED systems \cite{Johnson2010}. 



During the mapping process, the bosonic state $\ket{0}$ ($\ket{L}$) is mapped to the ancilla state $\ket{g}$ ($\ket{e}$), while all intermediate Fock states $\ket{0<n<L}$ are mapped to the ancilla state $\ket{e}$. Experimentally, this mapping can be realized by initializing the ancilla in the ground state, then utilizing the dispersive coupling to apply a collection of selective pulses \cite{Schuster2007, Johnson2010, Krastanov2015, Heeres2015, Reagor2016} at frequencies $(\omega_{ge} - k\chi)$ for $k = 0, 1, \ldots L$, where $\omega_{ge}$ is the bare frequency of the ancilla qubit. These pulses, which can be applied simultaneously, flip the ancilla to the excited state only if the bosonic mode state is $\ket{1}$, $\ket{2}$, ..., or $\ket{L}$. As a simpler alternative, a single selective pulse can be applied at $\omega_{ge}$ to flip the qubit conditioned on whether the bosonic mode is in $\ket{0}$. The only difference between this latter procedure and the mapping in~\hyperref[fig:setup]{Fig.~\ref{fig:setup}(b)} is that the roles of the ancilla states are reversed---a trivial change in bookkeeping.


Because readouts are frequently limited by qubit lifetime, we consider a bosonic mode that is subject to spontaneous relaxation. Specifically, the decay rate of a Fock state $\ket{n}$ to $\ket{n-1}$ is given by $n \kappa_\downarrow$, where the factor of $n$ is due to bosonic enhancement. Transitions between non-adjacent Fock states are suppressed by selection rules, and excitations will be considered later in \hyperref[decayexcite-twolevel]{Sec.~\ref{decayexcite-twolevel}}. 

As a figure of merit for this readout scheme, the readout fidelity $\mathcal{F}$ is defined as~\cite{Gambetta2007,Walter2017} 
\begin{equation}
\mathcal{F} = 1 - P(0_B|1_B) - P(1_B|0_B),
\label{eqn:fidelity}
\end{equation}
where $P(i|j)$ is the probability of the level-2 readout yielding $i$ when the initial state of the bosonic qubit was $j$, for $i,j \in \{ \ket{0}_B,\ket{1}_B \}$. $\mathcal{F}$ varies continuously from 0, for readouts which yield no information about the initial state, to 1, for perfect readouts. In the robust readout scheme, both $P(0_B|1_B)$ and $P(1_B|0_B)$ are suppressed by increasing $L$ and $N$, as is shown quantitatively in the following sections. 

Finally, to make the following analysis more concrete, in \hyperref[fig:setup]{Fig.~\ref{fig:setup}(c)} we show an example of a real system where the robust readout scheme can be applied---a circuit QED system. In this system, a microwave cavity mode (the bosonic mode) dispersively couples to a transmon qubit (the ancilla), and this coupling can be used to perform repeated QND measurements of the cavity state~\cite{Johnson2010, Peaudecerf2014,Sun2014,Ofek2016}. For a qubit stored in the cavity mode with a suitable encoding (e.g.~with the Fock, cat, or binomial codes), it will be shown that contributions to readout infidelity from cavity decay, mapping errors, and transmon readout errors can all be suppressed to higher order with this scheme.


\subsection{Discrete model of the robust readout scheme} 
\label{decay-twolevelB}
\begin{figure}[htbp]
\begin{center}
\includegraphics[width=\columnwidth]{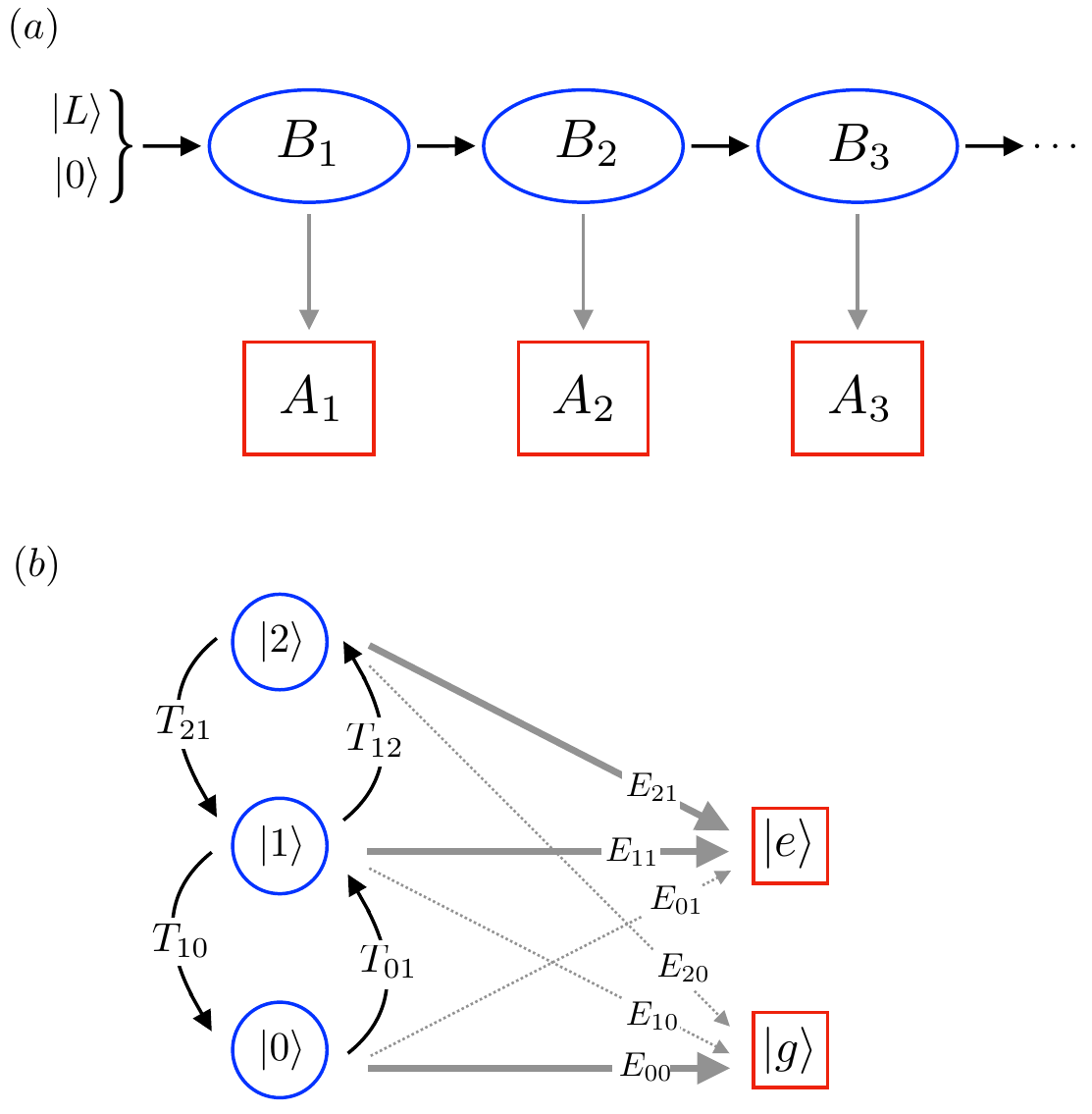}
\caption{Hidden Markov Model for the robust readout scheme. (a) Markov chain and emissions. At each step of the HMM the bosonic system transitions to a new state and releases an emission. Here, $B_n$ denotes the bosonic mode state after step $n$, and $A_n$ denotes the $n^{th}$ ancilla measurement outcome. (b) Transition and emission probabilities. Transitions and emissions are shown diagrammatically for the case $L=2$, where the matrix elements along the arrows are the associated probabilities. Bolded arrows indicate the intended mappings. }
\label{fig:HMM}
\end{center}
\end{figure}

A Hidden Markov Model (HMM) is used to model the robust readout scheme of \hyperref[fig:setup]{Fig.~\ref{fig:setup}}. A HMM is a Markov chain where, instead of being able to observe a system's state directly, the only information about the system is provided by a series of noisy emissions. HMMs have been previously used as effective models of qubit readout \cite{Dreau2013,Gammelmark2013,Ng2014,Wolk2015}. In our case, a discrete model (\hyperref[fig:setup]{Fig.~\ref{fig:HMM}}) is used where each level-1 readout is modeled as a possible transition, representing the bosonic qubit's decay, followed by a noisy emission, representing the mapping and the readout of the ancilla. 

The model is parameterized by transition probabilities $T_{ij}$ and emission probabilities $E_{ij}$. The transition probability $T_{ij}$ is defined to be the probability that the bosonic state $\ket{i}$ transitions to $\ket{j}$ during a single level-1 readout, with $i, j \in \{0, 1, \ldots, L \}$. The emission probability $E_{ij}$ is the probability that the bosonic system, having transitioned to state $\ket{i}$, with $i,\in \{0, 1, \ldots, L \}$, is read out as ancilla state $\ket{g}$ for $j=0$, or $\ket{e}$ for $j=1$. The emission probabilities are defined in terms of the probability $\delta$ that an error occurs during the mapping and readout processes which causes the ancilla readout to be misleading 
\begin{equation}
    E_{ij}= 
\begin{cases}
    1-\delta,& \text{if } i=j=0 \text{ or } i>0,j=1,\\
    \delta,& \text{otherwise. }
\end{cases}
\end{equation}
In cases where different Fock states have different probabilities of producing misleading ancilla readouts, taking $\delta$ to be the largest of these probabilities will yield a conservative estimate of readout fidelity.

Explicit expressions for transition probabilities $T_{ij}$ are derived from the bosonic decay rates. Consider a population of quantum harmonic oscillators, with $p_i(t)$ of the oscillators in Fock state $\ket{i}$ at time $t$. The system of differential equations describing the time evolution of the populations is
\begin{equation}
\dot{p}_i(t) = \sum_{j=0}^L (K_\downarrow)_{ij}\,p_j(t),
\end{equation}
where $(K_\downarrow)_{ij}$ is the transition rate from state $\ket{j}$ to $\ket{i}$. For bosonic systems,
\begin{equation}
(K_\downarrow)_{ij} =  \begin{cases} 
      -j\,\kappa_{\downarrow}, & i = j \\
       \,\,\,\,\,j\, \kappa_{\downarrow}, &i = j-1\\
       \,\,\,\,\,0, & \text{otherwise}.
   \end{cases}
\end{equation}
This system has the solution $\mathbf{p}(t) = e^{K_\downarrow t}\,\mathbf{p}(0)$.
The transition probabilities for a level-1 readout taking time $\tau$ are thus obtained by explicitly computing the matrix elements of $e^{K_\downarrow \tau}$,
\begin{equation}
T_{ij}(\tau)=(e^{K_\downarrow \tau})_{ji} = {{i}\choose{j}}\left(e^{\kappa_\downarrow \tau}-1\right)^{i-j} e^{-i\kappa_{\downarrow}\tau}.
\end{equation}

As an aside, we note that both $\tau$ and $\delta$ can depend implicitly on the strength of the dispersive coupling $\chi$. For example, larger coupling strengths can enable faster or more selective pulses. The values of $\tau$ and $\delta$ given in \hyperref[fig:setup]{Fig.~\ref{fig:setup}(c)} are estimated from the given $\chi$ value based on such considerations. In order to keep the following discussion general, however, we do not assume a particular functional dependence of either of these parameters on $\chi$.

To provide intuition as to why increasing the number of levels $L$ can improve the readout fidelity, we calculate the expected value of the time $\tau_0$ which it takes initial state $\ket{L}$ to decay to $\ket{0}$,
\begin{align}
\left< \tau_0\right> = \int_0^\infty  d\tau\, \tau\,  \frac{d\, }{d\tau} T_{L0}(\tau)  =\frac{1}{\kappa_\downarrow} \sum_{n=1}^L \frac{1}{n}.
\end{align}
Because $\tau_0$ grows with $L$, so too does the effective signal lifetime, thereby improving readout fidelity. Indeed, the effective lifetime diverges with $L$, though there are diminishing returns in using higher levels because the divergence is only logarithmic.  Interestingly, it should be noted that using higher-level encodings can improve readout fidelity even in the absence of an increase in effective signal lifetime \cite{DAnjou2017}.

\subsection{Readout infidelity in the discrete model}

Using the HMM, we calculate the infidelity $1-\mathcal{F}$ of the robust readout scheme in terms of the ``experimental'' parameters $\delta$ and $\kappa_{\downarrow}\tau$. This infidelity depends on how the level-2 measurement outcomes are determined. We consider two approaches: simple majority voting and a maximum likelihood estimate (MLE). 

In majority voting, each level-2 measurement outcome is determined by tallying the $N$ level-1 measurement outcomes, with ancilla readouts of $\ket{g}$ ($\ket{e}$) counted as votes for initial state $\ket{0}$ ($\ket{L}$). In the MLE, which is the statistically optimal approach, the known values of the transition and emission matrix elements are used to calculate which initial state was more likely to have produced a series of observed ancilla readouts. Explicitly, the likelihood $\lambda_{\mathbf{a}}(i)$ that a discrete set of ancilla readouts $a_n \in \{g,e \}$, for $n \in \{1, \ldots, N \}$, was produced with initial state $\ket{i}$ is 
\begin{equation}
\lambda_{\mathbf{a}}(i) = \sum_{j_1,\ldots,j_N} T_{i,j_1} E_{j_1,a_1} \ldots T_{j_{N-1},j_N}E_{j_N,a_N},
\end{equation}
which is efficiently calculable in $O(NL^2)$ operations \cite{Press2007}. 
The outcome of a level-2 measurement is then decided by determining which of the two initial states was more likely to have produced the emissions, i.e.~by comparing $\lambda_{\mathbf{a}}(0)$ and $\lambda_{\mathbf{a}}(L)$.  

For both majority voting and the MLE classification strategies, the infidelity is given exactly as a function of the likelihoods
\begin{align}
1- \mathcal{F} &= \sum_{\mathbf{a} \in \mathcal{A}_0 }\lambda_{\mathbf{a}}(L) + \sum_{\mathbf{a'} \in \mathcal{A}_L }\lambda_{\mathbf{a'}}(0).
\end{align}
where $\mathcal{A}_0$ ($\mathcal{A}_L$) is the set of ancilla readout vectors $\mathbf{a}$ which are classified as initial state $\ket{0}$ ($\ket{L}$). Whether a given $\mathbf{a}$ falls in either $\mathcal{A}_0$ or $\mathcal{A}_L$ depends on the classification strategy. By definition, the MLE chooses the sets $\mathcal{A}_0$ and $\mathcal{A}_L$ to be those which minimize the infidelity. 

\begin{figure}[htbp]
\begin{center}
\includegraphics[width=\columnwidth]{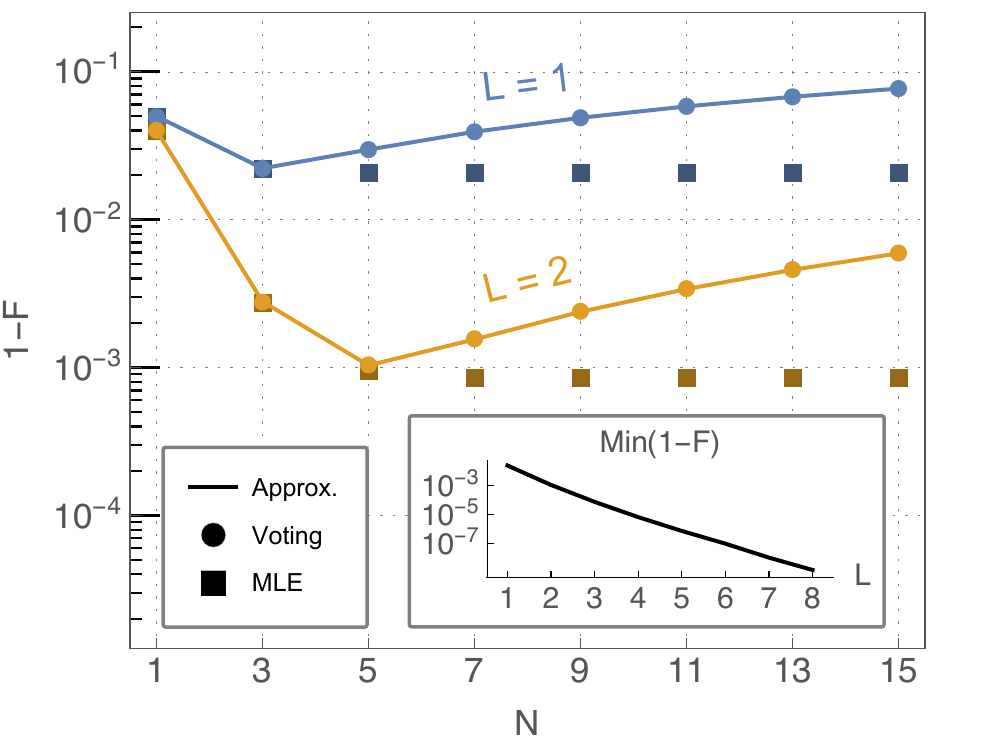}
\caption{Infidelity of the robust readout scheme. The infidelity is plotted as a function of the number of measurements $N$ for $L=1$ and 2, with the parameter choices $\delta = 2\%$ and $\kappa_{\downarrow}\tau = 1\%$. The solid lines denote the approximate majority voting infidelity~(\ref{eqn:approx}), while circles and squares respectively denote exact calculations for the majority voting and MLE schemes.  Inset: the minimum attainable infidelity is plotted as a function of $L$.}
\label{fig:decay}
\end{center}
\end{figure}

Plots of the infidelity as a function of $N$ are shown in \hyperref[fig:decay]{Fig.~\ref{fig:decay}} for both majority voting and the MLE. The values of $\kappa\tau$ and $\delta$ used in the figure are the same as those given for the circuit QED system in \hyperref[fig:setup]{Fig.~\ref{fig:setup}(c)}, so the infidelities shown in the main panel are realistically attainable. Notably, the minimum infidelity attained by both majority voting and the MLE decreases by over an order of magnitude as $L$ increases from 1 to 2.  Indeed, the inset shows that increasing $L$ can lead to multiple orders of magnitude improvement. It is also clear that the MLE can dramatically outperform majority voting as $N$ increases. This discrepancy is due to decays: majority voting weights all votes equally, even those that are recorded long after initial state $\ket{L}$ is likely to have decayed to $\ket{0}$. The minimal infidelities attained by the two methods, however, are not significantly different, meaning that simple majority voting is a near-optimal strategy until decays begin to play a significant role.

To compute the exact infidelity, it is necessary to enumerate all possible combinations of $N$ level-1 readouts and to compute the likelihoods of each, a computation which takes $O(NL^2\times2^N)$ operations.  To provide a more accessible means of quickly estimating the readout infidelity, and to elucidate its scaling, we derive a simple approximation for the infidelity in the majority voting scheme. The approximation depends on a small number of general experimental parameters: the level-1 readout error probability $\delta$, the decay rate of the bosonic system $\kappa_{\downarrow}$, and the level-1 readout time~$\tau$. 

There are two dominant processes which are most likely to fool the majority voting. The first is sufficiently quick decay of the initial state $\ket{L}$ to $\ket{0}$, but with no level-1 readout errors occurring. The second is a sufficient number of level-1 readout errors occurring so as to fool the voting, but with no decays occurring. All other processes which fool the voting, such as combinations of decays and level-1 readout errors, have probabilities that are higher order in the parameters $\delta$ or $\kappa_\downarrow \tau$. We approximate the probabilities of incorrectly identifying initial states by neglecting the contributions of these higher-order processes,
\begin{subequations}
\begin{align}
 P(0|L) &\approx T_{LL}\left(N\tau\right) \times \sum_{k = \ceil{N/2}}^L {{N}\choose{k}} \delta^k (1-\delta)^{N-k} \nonumber \\[0.1cm]
 &+  T_{L0}\left(\ceil{N/2}\tau\right)\times(1-\delta)^N \\[0.1cm]
 P(L|0) &\approx  \sum_{k = \ceil{N/2}}^L {{N}\choose{k}} \delta^k (1-\delta)^{N-k},
\end{align}
\end{subequations}
where $\ceil{\cdot}$ denotes the ceiling function.
Expanding to lowest order in $\delta$ and $\kappa_\downarrow \tau$ gives
\begin{align}
\label{eqn:approx}
1-\mathcal{F} & =  P(0|L) + P(L|0) \nonumber \\[0.2cm]
&\approx \, 2{{N}\choose{\ceil{N/2}}} \delta^{\ceil{N/2}} + \left( \ceil{N/2} \kappa_\downarrow \tau\right)^{L} .
\end{align}
This approximation is valid when both $N\delta \ll 1$ and $N\kappa_{\downarrow} \tau\ll1$ so that higher order terms can be neglected. This approximation is plotted along with the exact result in \hyperref[fig:decay]{Fig.~\ref{fig:decay}}, where the two agree well because the approximation is valid in the regime shown. 

\hyperref[eqn:approx]{Eqn.~\ref{eqn:approx}} elucidates the benefit of combining robust encoding with repeated measurement. In two-level systems, such as trapped ions, the fidelity is limited by $\kappa_\downarrow\tau$ because $L = 1$ is fixed. On the other hand, in multi-level systems where repetitive QND readouts are not possible, the fidelity is limited by $\delta$ because $N=1$ is fixed. For bosonic systems in the dispersive coupling regime, however, one has the freedom to increase both $L$ and $N$. Thus, both terms contributing to the infidelity are suppressed to higher order, and readout is no longer theoretically limited by either individual measurement errors or relaxation. This is the strength of the robust readout scheme.


\section{Robust readout with both relaxation and heating}
\label{decayexcite-twolevel}
We now consider the case where the bosonic mode is subject to heating, defined here as a nonzero excitation rate $\kappa_\uparrow$. Without modification, the readout fidelity of the above scheme would be limited by the probability of the initial state $\ket{0}$ spontaneously exciting to $\ket{1}$, a process which is first order in $\kappa_\uparrow\tau$. In this section, we generalize the scheme so that contributions to the infidelity from heating are also suppressed to higher orders.

\begin{figure}[htbp]
\begin{center}
\includegraphics[width=\columnwidth]{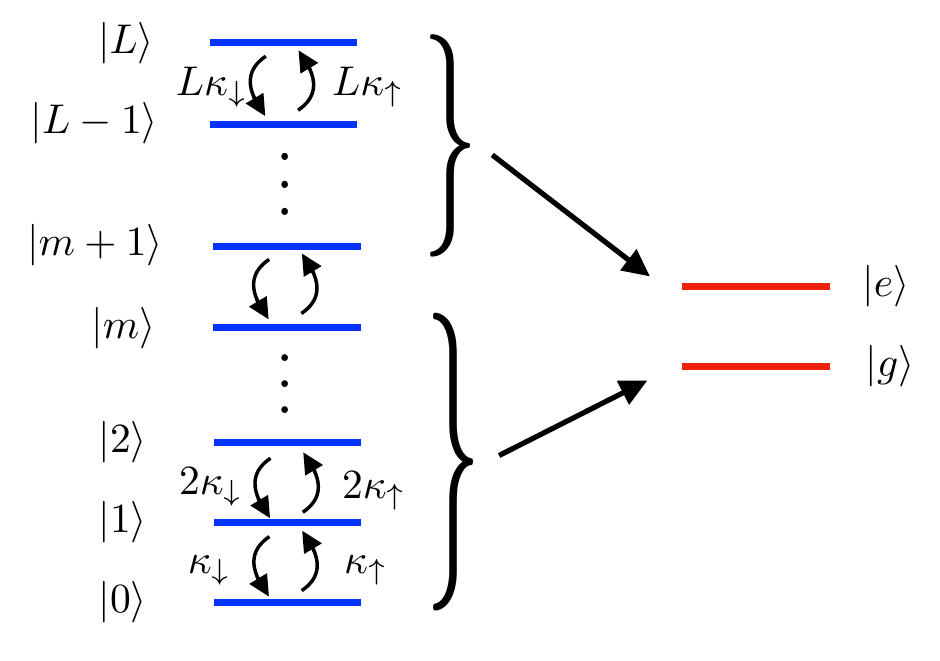}
\caption{Robust readout scheme for relaxation and heating. Decays and excitations occur between adjacent Fock states with rates proportional to the excitation number. All Fock states $\ket{n>m}$ are mapped to the excited state of the two-level ancilla. }
\label{fig:decayexcite_setup}
\end{center}
\end{figure}

The modified readout scheme is shown in \hyperref[fig:decayexcite_setup]{Fig.~\ref{fig:decayexcite_setup}}, where the excitation rate\footnote{In order to study the fidelity with a finite HMM, we truncate the Hilbert space to the first $L+1$ Fock states, taking the heating rate from $\ket{L}$ to $\ket{L+1}$ to be 0. It is safe to neglect the additional levels when $\kappa_{\uparrow}\tau \ll \kappa_{\downarrow}\tau \ll 1$.} between the adjacent Fock states $\ket{n}$ and $\ket{n+1}$ is $n\kappa_{\uparrow}$. 
To account for this heating, we define a threshold state $\ket{m}$ such that the mapping from the bosonic mode to the ancilla is
\begin{equation}
\ket{n} \rightarrow  
	\begin{cases} 
        \ket{g}, & n\leq m \\
       	\ket{e}, &n>m .
   \end{cases}
\end{equation}
This mapping can be implemented by initializing the ancilla in the ground state, then applying selective pulses at frequencies $(\omega_{ge} - k\chi)$ for $k = m+1,m+2, \ldots, L$. These pulses flip the ancilla from $\ket{g}$ to $\ket{e}$ only if the bosonic mode state is $\ket{m+1}$, $\ket{m+2}$, ..., or $\ket{L}$. The level-1 readouts are then described by the measurement operators
\begin{align}
\hat{M}_0 &= \,\,\, \sum_{k=0}^m\,\,\,\, \ket{k}\bra{k} \nonumber \\
\hat{M}_1 &= \sum_{k=m+1}^L \ket{k}\bra{k} .
\end{align}
For $m>0$, the contribution to the infidelity from heating of the initial state $\ket{0}$ will thus be suppressed to higher order in $\kappa_\uparrow$ because multiple excitations are required for $\ket{0}$ to heat to a state which is mapped to ancilla state $\ket{e}$.

As in the previous section, this scheme is quantitatively analyzed with a HMM. The emission probabilities $E_{ij}$ are similarly defined in terms of the level-1 readout error probability $\delta$ as
\begin{equation}
    E_{ij}= 
\begin{cases}
    1-\delta,& \text{if } i\leq m,j=0 \text{ or } i>m,j=1,\\
    \delta,& \text{otherwise.}
\end{cases}
\end{equation}
The transition probabilities $T_{ij}$ are calculated as functions of the decay and excitation rates. The system of differential equations describing the time-evolution of the Fock state populations is
\begin{equation}
\dot{p}_i(t) = \sum_{j=0}^L (K_\downarrow +K_\uparrow)_{ij}\,p_j(t),
\end{equation}
where $K_\uparrow$ has matrix elements
\begin{equation}
(K_\uparrow)_{ij} =  \begin{cases} 
      -(j+1)\,\kappa_{\uparrow}, & i = j < L \\
       \,\,\,\,\,(j+1)\, \kappa_{\uparrow}, &i = j+1 \\
       \,\,\,\,\,0, & \text{otherwise}.
   \end{cases}
\end{equation}
The transition probabilities are then given as a function of the level-1 readout time $\tau$,
\begin{equation}
T_{ij}(\tau)=\left[e^{( K_{\downarrow}+ K_{\uparrow})\tau}\right]_{ji}.
\end{equation}

Exact calculations of the infidelity proceed as in the previous section. We also approximate the infidelity by again considering only the dominant error processes, now including the probability that initial state $\ket{0}$ heats to $\ket{m+1}$, with no level-1 readout errors occurring. With this additional term, the level-2 readout error probabilities are approximately given by
\begin{subequations}
\begin{align}
 P(0|L) &\approx T_{LL}(N  \tau) \sum_{k = \ceil{N/2}}^L {{N}\choose{k}} \delta^k (1-\delta)^{N-k} \nonumber \\[0.1cm]
 &+ (1-\delta)^N\left(e^{K_\downarrow \ceil{N/2}\tau}\right)_{m,L} \\[0.1cm]
 P(L|0) &\approx T_{00}(N\tau) \sum_{k = \ceil{N/2}}^L {{N}\choose{k}} \delta^k (1-\delta)^{N-k}\nonumber \\[0.1cm]
 &+ (1-\delta)^N\left(e^{K_\uparrow \ceil{N/2}\tau}\right)_{m+1,L}.
\end{align}
\end{subequations}
To lowest order in $\delta$, $\kappa_\downarrow \tau$, and $\kappa_\uparrow \tau$, the infidelity is
\begin{align}
\label{heatinfidelity}
1-\mathcal{F} &\approx  {{L}\choose{m}}\left( \left\lceil \frac{N}{2}\right\rceil \kappa_\downarrow \tau\right)^{L-m}
  + \left(\left\lceil \frac{N}{2}\right\rceil \kappa_\uparrow \tau\right)^{m+1} \nonumber \\[0.1cm]
  &+2{{N}\choose{\ceil{N/2}}} \delta^{\ceil{N/2}} .
\end{align}
It is clear that, within this approximation, all contributions to the infidelity are suppressed to higher orders in $\kappa_\downarrow\tau$, $\kappa_\uparrow\tau $, and $\delta$, by increasing $L$, $m$, and $N$, respectively.

\begin{figure}[htbp]
\begin{center}
\includegraphics[width=\columnwidth]{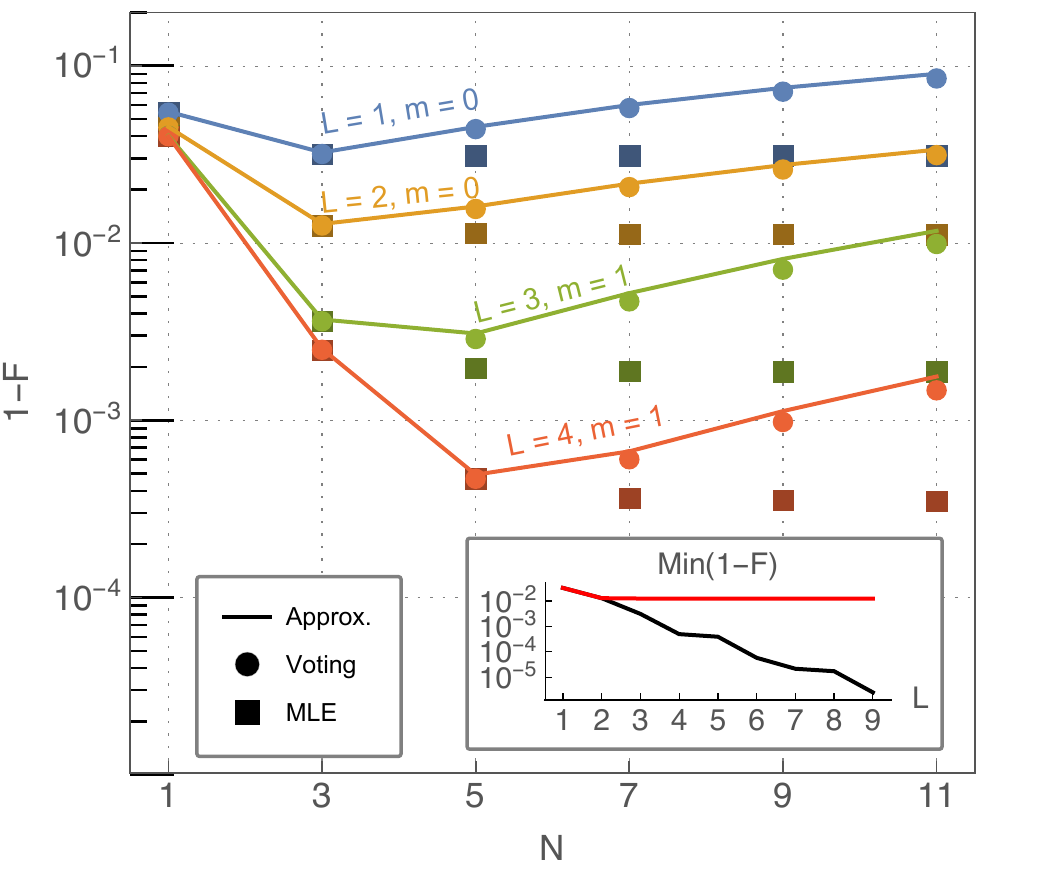}
\caption{
Infidelity of the robust readout scheme with both relaxation and heating. The infidelity is plotted as a function of the number of measurements $N$ for $L=1,2$ and 3, with the parameter choices $\delta = 2\%$, $\kappa_{\downarrow}\tau = 1\%$, and $\kappa_{\uparrow}\tau= 0.5\%$. Inset: the minimum attainable infidelity is plotted as a function of $L$. For $m=0$ (red) the infidelity asymptotes to a finite value, but for optimal $m$ (black) it continues to decrease. }
\label{fig:decayexciteinset}
\end{center}
\end{figure}

Plots of the infidelity with both majority voting and the MLE are shown in \hyperref[fig:decayexciteinset]{Fig.~\ref{fig:decayexciteinset}}. Though the heating rates $\kappa_\uparrow$ of physical systems are typically much smaller than the decay rate $\kappa_\downarrow$ (e.g.~\cite{Chen2016}), the two are chosen to be comparable in the plot so that the importance of the threshold state is apparent. For the parameters shown in the figure, $m=0$ is the optimal choice of the threshold for $L\leq2$, but at $L=3$ the optimal choice is $m=1$. In the inset, the minimum majority voting infidelity is plotted as a function of $L$ for both fixed $m=0$ (red) and the optimal choice of $m$ (black). It is clear that without increasing $m$ the readout infidelity is limited by the first-order heating process, but when $m$ is allowed to increase it is again possible to improve readout fidelity by orders of magnitude. We also note that here again the optimal MLE and majority voting infidelities do not differ significantly.



\section{Robust readout with a multi-level ancilla}
\label{decayexcite-manylevel}

There exist experimental systems where a bosonic mode can be dispersively coupled to an ancilla with more than two levels. Circuit QED systems provide one example; the higher excited states of a superconducting transmon qubit have been populated and measured in experiment \cite{Bianchetti2010,Peterer2015}. We now consider a version of the robust readout scheme applicable to such systems and show that the use of a multi-level ancilla can lead to significant improvements in readout fidelity when the MLE is used. 

\begin{figure}[htbp]
\begin{center}
\includegraphics[width=\columnwidth]{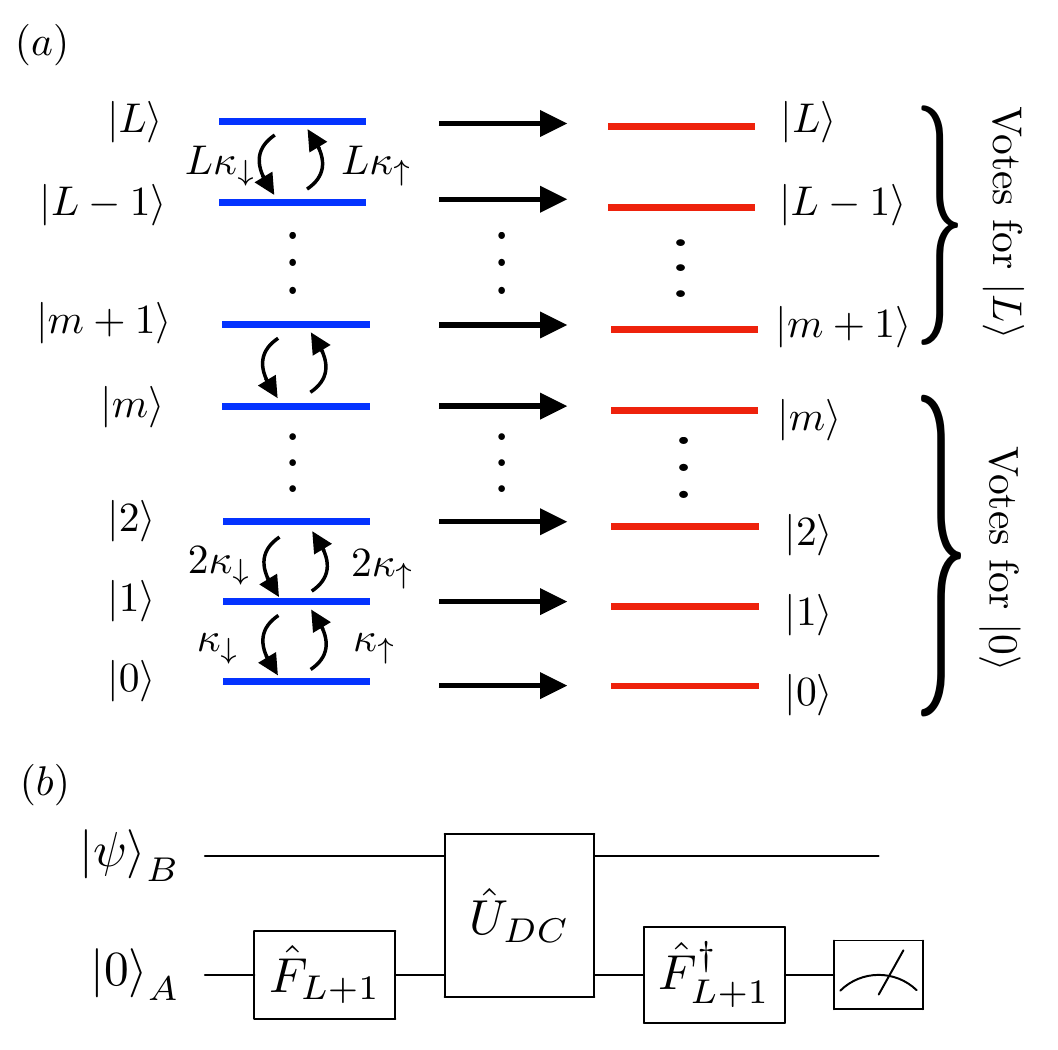}
\caption{Robust readout scheme for multi-level ancilla. (a) Schematic description. Both decays and excitations occur between adjacent bosonic mode Fock states. Each Fock state is mapped to a unqiue ancilla state. In the majority voting all ancilla readouts of $\ket{n>m}$ are counted as votes for $\ket{L}$, while readouts of $\ket{n\leq m}$ are counted as votes for $\ket{0}$. (b) Mapping circuit. Fourier gates on the ancilla, in combination with evolution under the dispersive coupling, implement the mapping used in the QND measurement. }
\label{fig:decayexciteL_setup}
\end{center}
\end{figure}

The readout scheme for this case is shown in \hyperref[fig:decayexciteL_setup]{Fig.~\ref{fig:decayexciteL_setup}}. As before, nonzero decay and excitation rates are assumed, but in this case the level-1 measurement operators are 
\begin{equation}
\{\hat{M}_k = \ket{k}\bra{k}, \text{ for $k=0,1,\ldots,L$}\}.
\end{equation}
The threshold state $\ket{m}$ is used only to determine which of the $L+1$ possible ancilla state readouts are counted as votes for initial bosonic state $\ket{0}$ or $\ket{L}$ in the majority voting scheme. It plays no role in the MLE.

A circuit that uses the dispersive coupling to implement the mapping from the bosonic mode to the ancilla is shown in \hyperref[fig:decayexciteL_setup]{Fig.~\ref{fig:decayexciteL_setup}(b)} \cite{Li2017}.  The ancilla is initialized in the ground state, and a Fourier gate $\hat{F}_{L+1}$ maps this state to an even superposition of the first $L+1$ Fock states. For a bosonic mode dispersively coupled to an $(L+1)$-level ancilla, the coupling Hamiltonian is 
\begin{equation}
\hat{H}_{DC}/\hbar = -\sum_{j=0}^{L} j\, \chi \ket{j}\bra{j}\hat{a}^\dagger \hat{a},
\end{equation}
 where $\ket{j}$ are the ancilla states. The bosonic mode and ancilla are allowed to evolve under this coupling for a time $t = 2\pi/(L+1)\chi$, implementing the unitary 
\begin{equation}
\hat{U}_{DC} = e^{i \frac{2\pi j}{L+1}  \ket{j}\bra{j}\hat{a}^\dagger \hat{a}  },
\end{equation}
after which the application of the gate $\hat{F}_{L+1}^\dagger$ completes the mapping of the bosonic mode's excitation number onto the ancilla. With this mapping, the measurement procedure is QND because the measurement operators $\hat{M}_k$ commute with the dispersive coupling.  



As a practical matter, we note that, since the number of excitations in the bosonic mode is not known \textit{a priori}, the dispersive coupling causes an unknown shift of the ancilla transition frequencies. However, this unknown frequency shift does not pose a barrier to implementing the Fourier gates in \hyperref[fig:decayexciteL_setup]{Fig.~\ref{fig:decayexciteL_setup}(b)}. If we can drive the ancilla with strength $\Omega$ much larger the dispersive coupling $\chi$, the standard control pulse has a small error decreasing with the driving strength as $(L\chi/\Omega)^2$. Moreover, dispersive coupling induced ancilla gate errors can be further suppressed to even higher order using composite pulses~\cite{Vandersypen2005} or numerically optimized control pulses~\cite{Khaneja2005, de_Fouquieres2011}.

The HMM transition probabilities $T_{ij}$ are the same as in the previous section, but it is necessary to redefine the emission probabilities $E_{ij}$ to incorporate the $L+1$ possible ancilla readouts. We define the emission matrix elements
\begin{align}
\label{emissionelements}
    E_{ij}= 
\begin{cases}
    (1-\delta),& \text{for } i = j  \\
    \delta/L,& \text{otherwise. }
\end{cases}
\end{align}
This choice\footnote{Note that with this definition $\delta$ is no longer the probability of obtaining a misleading readout. As a result, expressions involving $\delta$ in this section are not directly comparable to those in previous sections.} is made so that $\delta$ remains an easily measurable parameter: given the ability to reliably prepare an initial Fock state, $(1-\delta)$ is measurable as the probability that the state is correctly read out as the corresponding ancilla state. 

\begin{figure}[bp]
\begin{center}
\includegraphics[width=\columnwidth]{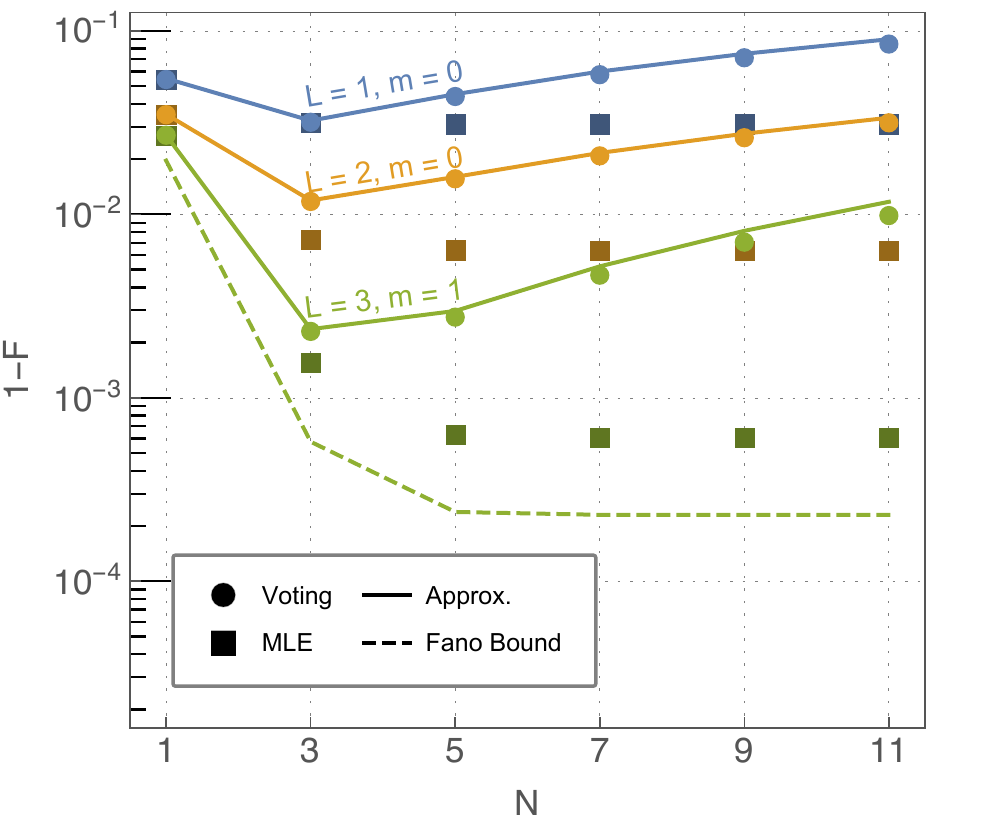}
\caption{ Infidelity of the robust readout scheme with multi-level ancilla. The infidelity is plotted with the parameter choices $\delta = 2\%$, $\kappa_{\downarrow}\tau = 1\%$, and $\kappa_{\uparrow}\tau= 0.5\%$. The dashed line is a lower bound on the fidelity determined through information-theoretic considerations (see \hyperref[InfoTheory]{Sec.~V}). }
\label{fig:decayexciteL}
\end{center}
\end{figure}
As before, the infidelity of the level-2 readout for both the majority voting and MLE is exactly calculable with the HMM. We also approximate the infidelity for the majority voting scheme:
\begin{align}
&1-\mathcal{F} \approx  {{L}\choose{m}}\left( \left\lceil \frac{N}{2}\right\rceil \kappa_\downarrow \tau\right)^{L-m}
+ \left(\left\lceil \frac{N}{2}\right\rceil \kappa_\uparrow \tau\right)^{m+1} \nonumber \\[0.1cm]
&+{{N}\choose{\ceil{N/2}}}\left[ \left(\frac{(m+1)}{L}\delta\right)^{\left\lceil \frac{N}{2}\right\rceil}+\left(\frac{(L-m)}{L}\delta\right)^{\left\lceil \frac{N}{2}\right\rceil} \right].
\end{align}
Representative infidelities are plotted in \hyperref[fig:decayexciteL]{Fig.~\ref{fig:decayexciteL}}. The most salient feature of the plot is the discrepancy between the minimum infidelities attained by the majority voting and the MLE. Whereas in the previous cases the two were not found to differ significantly, here the MLE is a clearly superior strategy. This discrepancy is due to the fact that the majority voting uses only binary information (votes for $\ket{0}$ or $\ket{L}$) to classify the $N$ level-1 outcomes. In contrast, the MLE can take any of the $L+1$ possible ancilla readouts as input and thus extracts more information from each level-1 readout. With this additional information, the MLE is able to more accurately determine the initial state. We further explore an information-theoretic description of the robust readout scheme in the next section.

\section{information-theoretic description }
\label{InfoTheory}
In this section, we consider the fidelity of the robust readout scheme from the perspective of classical information theory. The initial state of the bosonic mode constitutes one bit\footnote{In this section, all logarithms are base 2. } of information, and it is the goal of the robust readout scheme to extract as much of this information as possible. By quantifying the amount of information extracted, it is possible to place a general lower bound on the readout infidelity. 

We treat the initial state of the bosonic mode as a classical discrete random variable $B$ and suppose that initial states $\ket{0}$ and $\ket{L}$ are equally likely,
\begin{equation}
p_B(b) = \frac{1}{2},
\end{equation}
where $b \in \{0,L\}$ is a realization of $B$. 
Similarly, we treat the series of $N$ ancilla readouts as a discrete random variable $A$. The conditional probability distribution of $A$ given $B$ is given by the likelihood
\begin{align}
&p_{A|B}(\mathbf{a}|b) = \lambda_{\mathbf{a}}(b) \nonumber\\
= &\sum_{j_1, \ldots j_N} T_{bj_1}E_{j_1a_1}\ldots T_{j_{N-1}j_N}E_{j_N a_N},
\end{align}
where $\mathbf{a}$, an $N$-vector whose components are the ancilla measurement outcomes, is a realization of $A$. (For a two-level ancilla, $a_i \in \{0,1\}$, while for an $(L+1)$-level ancilla $a_i \in \{0,1,\ldots,L\}$.) We also calculate the remaining distributions in terms of the likelihoods: the joint probability distribution for $A$ and $B$,
\begin{equation}
p_{AB}(\mathbf{a},b) = \frac{1}{2}\lambda_{\mathbf{a}}(b);
\end{equation}
the marginal probability distribution for $A$,
\begin{align}
p_A(\mathbf{a}) =  \frac{\lambda_{\mathbf{a}}(0) + \lambda_{\mathbf{a}}(L) }{2};
\end{align}
and the conditional probability distribution of $B$ given $A$,
\begin{equation}
p_{B|A}(b|\mathbf{a}) =  \frac{\lambda_{\mathbf{a}}(b)}{\lambda_{\mathbf{a}}(0) + \lambda_{\mathbf{a}}(L)}.
\end{equation} 

The bosonic mode's initial state contains one bit of information, as quantified by the entropy $H$ of random variable $B$,
\begin{equation}
H(B) = -\sum_b p_B(b)\log(p_B(b)) = 1.
\end{equation}
The goal of the robust readout scheme is to indirectly extract as much of this information as possible through random variable $A$. The conditional entropy
\begin{equation}
H(B|A) = - \sum_{\mathbf{a},b}p_{AB}(\mathbf{a},b) \log(p_{B|A}(b|\mathbf{a})),
\end{equation}
quantifies the amount of uncertainty in $B$ given $A$, and it follows that the mutual information
\begin{equation}
I(A;B) = H(B) - H(B|A)
\end{equation}
quantifies the amount of information extracted through the robust readout procedure.

These quantities are used to bound the readout fidelity. Consider a classification process where one attempts to determine $B$ from $A$. Let $\hat{B}(A)$ be the guessed value of $B$. The probability of an incorrect assignment $P(\hat{B}(A) \neq B) \equiv p_e$  is related to the conditional entropy through \textit{Fano's inequality},
\begin{equation}
H(B|A) \leq H_2(p_e) + p_e \log(|\mathcal{B}|-1).
\end{equation}
Here, $\mathcal{B}$ is the support of random variable $B$, and $H_2$ is the binary entropy,
\begin{equation}
H_2(p_e) = -p_e \log(p_e) - (1-p_e) \log(1-p_e).
\end{equation}
Thus, Fano's inequality places a lower bound on the infidelity of the robust readout scheme $(1-\mathcal{F}) = 2\,p_e$, and the bound is calculable in terms of the relaxation and heating probabilities, $\kappa_\downarrow\tau$ and $\kappa_\uparrow\tau$, and the level-1 readout error probability $\delta$. 

This lower bound is shown in \hyperref[fig:decayexciteL]{Fig.~\ref{fig:decayexciteL}} for the case of $L=3$. This bound behaves similarly to the MLE, since the MLE is the optimal classification strategy. Despite the fact that the bound is not saturated, it is clear from the figure that classical information theory provides a reasonable alternative perspective from which the fidelity of the robust readout scheme can be understood. 

For completeness, we show why the MLE does not attain the bound. The bound is saturated only if $H(B|A) = H_2(p_e)$, since $|\mathcal{B}|=2$.  Equivalently, this condition may be written as $H(E|A) = H(E)$, where $E$ is the discrete random variable   
\begin{equation}
E  = \begin{cases}
1, &  \hat{B} \neq B\\
0, &  \hat{B} = B.
\end{cases}
\end{equation}
Qualitatively, $H(E|A) = H(E)$ holds when $A$ does not provide any information about whether a classification error will happen, i.e.~when classification errors are equally likely for all realizations of $A$. This property does not generally hold for the robust readout scheme since typically $P(0|L) \neq P(L|0)$. This is a consequence of the asymmetry between relaxation and heating rates, which enables one to be more confident in a correct classification for some sequences of ancilla readouts over others.

\section{Robust readout for bosonic encodings}
\label{Encodings}
Given a qubit stored in a bosonic mode as $\ket{\psi}_B = \alpha \ket{0}_B + \beta \ket{1}_B$, we have thus far only considered readout using the Fock state encoding
 \begin{align}
 \label{FockEncoding}
 \ket{0}_B &= \ket{0} \nonumber \\
 \ket{1}_B &= \ket{L}.
 \end{align}
This choice was made for simplicity---with this encoding the readout fidelity can be computed classically. While this error-detecting code could be useful in a concatenated architecture~\cite{Knill2005}, it may not be ideal for more general applications. Thus, in this section we consider alternate encodings. We develop a set of sufficient encoding criteria for the robust readout procedure to be applicable, show how these criteria are satisfied by cat codes and binomial codes, and approximate the majority voting readout fidelity for both encodings.

\subsection{Encoding criteria} 
For a qubit encoded in a lossy bosonic mode as $\ket{\psi}_B = \alpha \ket{0}_B + \beta \ket{1}_B$, 
we identify three encoding criteria that are sufficient for robust, ancilla-assisted readout in the $\{\ket{0}_B, \ket{1}_B \}$ basis.

\textit{Criterion 1:} Encodings must be robust against excitation loss so that a single loss error cannot destroy all information about the initial state. Explicitly, when subject to $k$ excitation losses, let the logical states $\ket{0}_B$ and $\ket{1}_B$ be respectively mapped to error states $\ket{E_0^{k}}$ and $\ket{E_1^{k}}$. 
The encoding is said to be robust against $d$ excitation losses if 
\begin{equation}
\left\langle E_0^{k} | E_1^{\ell} \right\rangle = 0,\, \text{ for } k \text{ and }\ell \in \{0,1,\ldots, d\},
\end{equation}
where $\ket{E_0^{0}}$ $(\ket{E_1^{0}})$ denotes $\ket{0}_B$ ($\ket{1}_B$).
For example, the Fock state encoding (\ref{FockEncoding}) is robust against $d= L-1$ excitation losses.
We note that this criterion is less stringent than the Knill-Laflamme conditions for quantum error correction \cite{Knill1997} because we only need to protect a bit of $\textit{classical}$ information.

\textit{Criterion 2:} The two logical states and their corresponding error states must be distinguishable through an ancilla readout procedure that is QND. For a projective measurement described by $\{ \hat M_k\}$ that is capable of distinguishing these states, the measurement is QND if
\begin{equation}
\left[\hat{H}(t), \hat{M_k} \right]   = 0\,  \text{ for all $k$},
\end{equation}
where $\hat H(t)$ is the Hamiltonian describing the readout procedure. The satisfaction of this criterion enables repeated readouts. As an example, a measurement described by the operators
\begin{align}
\hat{M}_0 &= \ket{0}_B \bra{0}_B + \ket{E_0^1} \bra{E_0^1} + \ldots + \ket{E_0^d} \bra{E_0^d} \nonumber \\
\hat{M}_1 &= \ket{1}_B \bra{1}_B + \ket{E_1^1} \bra{E_1^1} + \ldots + \ket{E_1^d} \bra{E_1^d},
\end{align}
is capable of distinguishing the logical states and their corresponding error states, and it is QND if both $\hat M_0$ and $\hat M_1$ commute with $\hat H(t)$.
For the two-level ancilla readout procedure of \hyperref[decay-twolevel]{Sec.~II}, the measurement operators (\ref{Eqn:decay_meas_ops}) commute with the dispersive coupling Hamiltonian, thereby satisfying this criterion.

\textit{Criterion 3:} Ancilla errors must not induce damaging changes in the bosonic mode's state. Let possible ancilla errors be described by a set of jump operators $\{\hat{J}_\ell\}$. For an ancilla error occurring at time $t$ during a level-1 readout, the evolution of the combined system is described by the operator
\begin{align}
\hat{J}_\ell '(t) =& \mathcal{T} e^{- \frac{i}{\hbar} \int_t^\tau \hat{H}(t') dt' }\,
 \hat{J}_\ell\, \mathcal{T}e^{- \frac{i}{\hbar} \int_0^t \hat{H}(t') dt'},
\end{align}
where $\mathcal{T}$ denotes time-ordering.
We must have
\begin{equation}
\left[\hat{J}_\ell '(t), \hat{M_k} \right] = 0, \text{ for all $k$ and $\ell$},
\end{equation}
so that ancilla jumps do not affect measurement outcomes by altering the bosonic mode state. 

More concretely, for a $d$-level ancilla we consider the possible ancilla errors
\begin{equation} 
\hat{J} \in \left\{ \ket{n}\bra{m}, \text{ for $n\neq m$ and $n,m \leq d$ }  \right\},
\end{equation}
corresponding to spontaneous transitions of the ancilla state. In the dispersive coupling regime, such jumps induce dephasing of the bosonic mode that can be modeled as applications of the operator $\hat{J}'\sim\hat{n}$ and its higher powers \cite{Michael2016,Reagor2016}. Therefore, we must have $[\hat{n},\hat{M}_k] = 0$ for this criterion to be satisfied, lest readout fidelity be limited by the probability of spontaneous ancilla transitions.


These three criteria are satisfied by the Fock state encoding (\ref{FockEncoding}).  We now show explicitly that the criteria are also satisfied by cat codes and binomial codes, and we approximate the fidelity of the robust readout scheme for both types of codes.


\subsection{Cat codes}

Cat codes \cite{Cochrane1999,Leghtas2013, Mirrahimi2014, Li2017} are quantum error correcting codes designed to protect against excitation loss. Quantum error correction with cat codes has recently reached the break-even point where the lifetime of encoded qubits exceeds the lifetimes of all constituent components \cite{Ofek2016}. The codewords are formed from equal superpositions of coherent states. Let the state $\ket{C_\alpha^{n}}$ be defined as a superposition of $2L$ coherent states evenly distributed around a circle in the bosonic mode's phase space
\begin{align}
\ket{C_\alpha^{n}} &= \frac{1}{2L\sqrt{N_\alpha^n}} \sum_{k=0}^{2L-1} e^{-ikn\pi/L} \ket{e^{ik\pi/L}\alpha},
\end{align}
where $N_\alpha^n$ is a normalization factor \cite{Li2017}.
These sates can be expressed in terms of Fock states as
\begin{align}
\ket{C_\alpha^{n}} &= 
\frac{1}{\sqrt{N_\alpha^n}}\sum_{m=0}^{\infty}\frac{e^{-|\alpha|^2/2}	\alpha^{n+2mL}}{\sqrt{(n+2mL)!}}\ket{n+2mL}_F
\end{align}
where the subscript $F$ is used in this section to distinguish Fock states from coherent states. It is important to note that $\ket{C_\alpha^{n}}$ is a superposition of Fock states which all have the same excitation number $n$ modulo $2L$. We define the logical states
\begin{align}
\ket{0}_B &= \ket{C_\alpha^L} \nonumber\\
\ket{1}_B &= \ket{C_\alpha^{2L}}.
\end{align}

\textit{Criterion 1.} After $k$ excitation loss events, the state $\ket{C_\alpha^{n}}$ is mapped to $\ket{C_\alpha^{n-k}}$.  The cat codes are robust against $L-1$ excitation loss events since
\begin{equation}
\left\langle C_\alpha^{L-k} | C_\alpha^{2L-\ell} \right\rangle = 0, \text{ for $k,\ell \leq L-1$}.
\end{equation} 

\textit{Criterion 2.} The cat code logical states and their corresponding error states can be distinguished by measurement of the excitation number modulo $2L$. This measurement can be described by the set of measurement operators $\{\hat{M}_k,  \text{ $k = 0, \ldots, 2L -1$}\}$, where
\begin{equation}
\label{Eqn:cat_ops}
\hat{M}_k = \sum_{m = 0}^\infty \ket{k + 2Lm }\bra{k + 2Lm }.
\end{equation}
This measurement can be implemented using the dispersive coupling $\hat{H}_{DC}$ with a procedure similar to the one shown in \hyperref[fig:decayexciteL_setup]{Fig.~\ref{fig:decayexciteL_setup}(b)}. Using Fourier gates on the $2L$-level ancilla, in combination with evolution under the dispersive coupling, implement the unitary
\begin{equation}
\hat{U} = \hat F_{2L}^\dagger e^{i \frac{2\pi j}{2L}\ket{j}\bra{j} \hat a ^\dagger \hat a}\hat F_{2L},
\end{equation}
which maps the bosonic mode's excitation number modulo $2L$ onto the ancilla. This measurement process is QND because 
$\left[\hat{H}_{DC}, \hat{M_k} \right] = 0$ for all $k$.

\textit{Criterion 3.} Spontaneous ancilla transitions during the readout process do not induce damaging changes in the bosonic mode's state because the measurement operators $\hat{M}_k$ commute with dephasing errors $\hat{n}$ for all $k$.


\textit{Fidelity.} To approximate the fidelity of the majority voting scheme we consider the two processes most likely to fool the voting: (1) sufficient level-1 readout errors with no excitation loss events, and (2) $L$ excitation loss events occurring sufficiently quickly with no level-1 readout errors. The probability of process (1) can be computed in terms of $\delta$, the probability of obtaining a misleading level-1 readout, as in the previous sections. To compute the probability of process (2), we first note that the Kraus operator-sum representation for the lossy bosonic channel \cite{Chuang1997} is
\begin{equation}
\mathcal{L}(\hat\rho) = \sum_{k=0}^\infty \hat A_k\, \hat \rho\, \hat A_k^\dagger,
\end{equation}
where 
\begin{equation}
\hat A_k = \sqrt{\frac{(1-e^{-\kappa_\downarrow t})^k}{k!}} e^{-\kappa_\downarrow t \hat n/2} \hat a^k
\end{equation}
is the Kraus operator corresponding to $k$ excitation losses. The probability of process (2) is the probability of initial state $\ket{C_\alpha ^n}$ suffering $L$ excitation loss events in a time $\ceil{N/2}\tau$, which is approximately given by 
\begin{align}
\left\langle \hat A_L^\dagger \hat A_L \right\rangle &\approx \frac{(\ceil{N/2}\kappa_\downarrow \tau)^L}{L!}
 \left\langle \hat a^{\dagger L} \hat a^L \right\rangle \nonumber \\
&\approx \frac{1}{L!} \left(|\alpha|^2\ceil{N/2}\kappa_\downarrow\tau \right)^L.
\end{align} 
To lowest order in $\delta$ and $\kappa_\downarrow\tau$, the cat code readout fidelity $\mathcal{F}_{\text{cat}}$ is thus given by
\begin{equation}
\mathcal{F}_{\text{cat}} \approx 1 - 2{{N}\choose{\ceil{N/2}}}\delta^{\ceil{N/2}} - \frac{2}{L!} \left(|\alpha|^2\ceil{N/2}\kappa_\downarrow\tau \right)^L.
\end{equation}
Within this approximation it is clear that both error terms are suppressed to higher order. The contribution from individual measurement infidelity is suppressed by increasing $N$, and the contribution from excitation loss is suppressed by increasing the number of coherent states comprising the cat state---analogous to increasing the excitation number used in the Fock state encoding.


\subsection{Binomial codes}
Binomial codes \cite{Michael2016} are a new class of quantum error correcting codes that can protect against excitation loss and gain errors as well as dephasing errors. The codewords are formed from superpositions of Fock states weighted with binomial coefficients
\begin{align}
\label{Eqn:binomial_codes}
\ket{0}_B &= \frac{1}{\sqrt{2^{M-1}}}\sum_{p \text{ even}}^{[0,M]}\sqrt{{M}\choose{p}}\ket{pL} \nonumber \\
\ket{1}_B &= \frac{1}{\sqrt{2^{M-1}}}\sum_{p \text{ odd}}^{[0,M]}\sqrt{{M}\choose{p}}\ket{pL},
\end{align}
where $M$ and $L$ are positive integers, and the range of the index $p$ is from $0$ to $M$.

\textit{Criterion 1.}  The error state $\ket{E_0^k}$ is a superposition of Fock states with excitation number $L-k$ mod $2L$, while error state $\ket{E_1^\ell}$ is a superposition with excitation number $2L-\ell$ mod $2L$. Therefore, the binomial codes are robust against $L-1$ excitation loss events since $\left\langle E_0^{k} | E_1^{\ell} \right\rangle = 0$ for $k$ and $\ell$ between 0 and $d$.

\textit{Criterion 2.} The binomial code logical states and corresponding error states can be distinguished by measuring the excitation number modulo $2L$. This measurement (\ref{Eqn:cat_ops}) is the same as that considered for cat codes, and it is QND by the same argument.

\textit{Criterion 3.}  Spontaneous ancilla transitions during the readout process do not induce damaging changes in the bosonic mode's state by the same argument as for cat codes.


\textit{Fidelity.} We approximate the fidelity of a majority voting scheme by considering the two processes most likely to fool the voting. The argument here proceeds analogously to the one given for cat codes, except that the probability of process (2) is different for binomial codes.
The probability that one of the initial states (\ref{Eqn:binomial_codes}) suffers $L$ excitation loss events in a time $\ceil{N/2}\tau$ is approximately given by
\begin{align}
\left\langle \hat A_L^\dagger \hat A_L \right\rangle &\approx \frac{(\ceil{N/2}\kappa_\downarrow \tau)^L}{L!} 
\left\langle \hat a^{\dagger L} \hat a^L \right\rangle \nonumber \\
&\approx \frac{1}{L!} \left( \frac{LM}{2} \ceil{N/2}\kappa_\downarrow\tau \right)^L.
\end{align} 
To lowest order in $\delta$ and $\kappa_\downarrow\tau$, the binomial code readout fidelity $\mathcal{F}_{\text{bin}}$ is then given by
\begin{equation}
\mathcal{F}_{\text{bin}} \approx 1 - 2{{N}\choose{\ceil{N/2}}}\delta^{\ceil{N/2}} - \frac{2}{L!}  \left( \frac{LM}{2} \ceil{N/2}\kappa_\downarrow\tau \right)^L.
\end{equation}
As with the cat codes, it is clear that both error terms are suppressed to higher orders.

\section{Conclusions}
We have shown how the combination of robust encoding and repeated QND measurements constitutes a powerful means of improving qubit readout fidelity. Robust encodings allow one to suppress contributions to the infidelity from relaxation, and repeated QND measurements allow one to suppress contributions from individual measurement infidelity. For bosonic systems in the dispersive coupling regime, these two techniques are simultaneously applicable. 
Strong dispersive couplings have already been experimentally demonstrated in circuit QED systems~\cite{Wallraff2004, Schuster2007,Boissonneault2009}, meaning the robust readout scheme can be readily applied, potentially yielding orders of magnitude improvement in readout fidelity. In principle, the scheme could also be applied to optomechanical~\cite{Jayich2008, Thompson2008}, nanomechanical~\cite{LaHaye2009, OConnell2010}, circuit quantum acoustodynamic~\cite{Manenti2017,Chu2017}, or quantum magnonics systems~\cite{Tabuchi2015, Tabuchi2016, Lachance-Quirion2017}. 

In this work we have not only studied the fidelity of the scheme for a simple Fock state encoding, but we have also provided general criteria that characterize other applicable encodings. We have shown that both cat codes and binomial codes can be read out robustly, thereby providing examples of quantum error correcting codes where the robust readout scheme is applicable. Ultra-high-fidelity logical state readout would be of great practical use in a number of applications where measurement fidelity is prioritized, including gate teleportation, entanglement purification, and modular quantum computation. 

\section{Acknowledgements}
We thank K.~Noh for helpful discussions. We acknowledge support from the ARL-CDQI, ARO (Grants No. W911NF-14-1-0011 and No. W911NF-14-1- 0563), ARO MURI (Grant No. W911NF-16-1-0349), NSF (Grants No. DMR-1609326 and No. DGE-1122492), AFOSR MURI (Grants No. FA9550-14-1-0052 and No. FA9550-15-1-0015), the Alfred P. Sloan Foundation (Grant No. BR2013-049), and the Packard Foundation (Grant No. 2013-39273).


\normalbaselines 
\bibliography{ReadoutReferences2} 

\end{document}